\documentclass[12pt]{article}
\usepackage{amsfonts}
\usepackage{amsmath}
\usepackage{amssymb}
\usepackage{graphicx}
\usepackage{color}
\usepackage[left=1cm,top=1cm,bottom=1cm,right=1cm,nohead,nofoot]{geometry}

\def \be {\begin{equation}}
\def \ee {\end{equation}}
\def \bea {\begin{eqnarray}}
\def \eea {\end{eqnarray}}
\def \nn {\nonumber}

\def \rr {\raise.35ex\hbox{\small $\prime$}\kern-.17em{\mbox{\large $\imath$}}}
\def \del {\partial}
\def \dels {\partial\kern-.6em /\kern.1em}
\def \As {{A\kern-.5em / \kern.5em}}
\def \Ds {D\kern-.7em / \kern.5em}
\def \a {\alpha}

\def \b {\beta}

\def \g {\gamma}

\def \d {\delta}
\def \eps {\epsilon}

\def \ks {k\kern-.5em /}
\def \ls {l\kern-.5em /}\def \lam {\lambda}

\def \II {I\hspace{-.1em}I\hspace{.1em}}

\def \IIA {\mbox{\II A\hspace{.2em}}}

\def \dd {\dot{\delta}}

\def \dm {\dot{\mu}}
\def \dn {\dot{\nu}}

\def \dlam {\dot{\lambda}}
\def \ds {\dot{\sigma}}
\def \dr {\dot{\rho}}

\reversemarginpar
\newcommand{\gnum}[1]{[#1]_g}
\newcommand{\hide}[1]{}

\begin{document}
\begin{titlepage}

\begin{center}

\hfill
\vskip .2in

\textbf{\LARGE
S-Duality for D3-Brane in \\ 
NS-NS and R-R Backgrounds
\vskip.5cm
}

\vskip .5in
{\large
Pei-Ming Ho$\,{}^{a,b,c,}$\footnote{e-mail address: pmho@phys.ntu.edu.tw} and
Chen-Te Ma$^{a,}$\footnote{e-mail address: yefgst@gmail.com}\\
\vskip 3mm
}
{\sl
${}^a$
Department of Physics and Center for Theoretical Sciences, \\
${}^b$
Center for Advanced Study in Theoretical Sciences, \\
${}^c$
National Center for Theoretical Sciences, \\
National Taiwan University, Taipei 10617, Taiwan,
R.O.C.}\\
\vskip 3mm
\vspace{60pt}
\end{center}
\begin{abstract}

We construct the low-energy effective field theory 
for a D3-brane in constant R-R 2-form potential background 
as the S-dual theory of a D3-brane in NS-NS $B$-field background.
Despite the non-Abelian algebra of the noncommutative $U(1)$ gauge symmetry,
the electromagnetic duality transformation can be carried out to all orders, 
and the dual Lagrangian is given in a compact form.
The gauge algebra is found to be a mixture 
of a deformed area-preserving diffeomorphism 
and the usual $U(1)$ gauge symmetry.

\end{abstract}

\end{titlepage}

\section{Introduction}
\label{1}


The effective theory for a D-brane in constant NS-NS $B$-field background
is a noncommutative gauge theory 
\cite{Chu-Ho,Schomerus:1999ug,Seiberg:1999vs}. 
By S-duality, 
one expects that the effective theory of a D3-brane 
in the background of constant R-R 2-form potential
to be the electromagnetic dual of 
a 4-dimensional noncommutative gauge theory.
In this paper,
we construct the D3-brane effective Lagrangian in R-R background
by carrying out the electromagnetic duality transformation 
for a 4-dimensional noncommutative $U(1)$ gauge theory.

For an Abelian gauge symmetry, 
the electromagnetic duality interchanges the field strength $F_{\mu\nu}$ 
and its Hodge-dual 
$\tilde{F}_{\mu\nu} \equiv \frac{1}{2} \epsilon_{\mu\nu\lambda\rho}F^{\lambda\rho}$.
With the field strength defined without introducing the gauge potential,
but required to satisfy the Bianchi identity,
the duality interchanges the field equation and the Bianchi identity
\be
\del_{\mu}F^{\mu\nu} = 0 
\quad \leftrightarrow \quad
\del_{\mu}\tilde{F}^{\mu\nu} = 0.
\ee
(For simplicity, 
let us focus on source-free configurations.)
A crucial feature is that 
the gauge potential does not appear explicitly 
in either equations.

Although S-duality continues to hold for multiple D3-branes,
the electromagnetic duality for non-Abelian gauge theory 
cannot be made as explicit as the Abelian case.
The reason is that 
the field equation and the Bianchi identity for the non-Abelian gauge theory
\be
D_{\mu}F^{\mu\nu} = 0, 
\qquad
D_{\mu}\tilde{F}^{\mu\nu} = 0
\ee
depend explicitly on the gauge potential $A$ through
the covariant derivative $D_{\mu} = \del_{\mu} + A_{\mu}$.
Interchanging $F$ with $\tilde{F}$ is not consistent with 
interchanging the two equations above because 
$A$ cannot remain the same when $F$ and $\tilde{F}$ are interchanged. 
As there is no formulation of the non-Abelian gauge theory 
that does not refer explicitly to the gauge potential, 
even for the source-free theory, 
the electromagnetic duality of a non-Abelian gauge theory 
has been mysterious \cite{Deser}.

To the lowest order approximation, 
the noncommutative gauge algebra is a non-Abelian algebra
defined by the Poisson bracket.
(The Moyal bracket defining the noncommutative algebra 
can be viewed as its deformation.)
The electromagnetic dual at the Poisson level 
is in fact already known.
It is derived from 
the effective theory of an M5-brane in large constant $C$-field background 
\cite{M51,M52,Ho:2009zt,Ho:2012dn}.
The non-Abelian gauge symmetry of this theory is 
characterized by the Nambu-Poisson bracket, 
hence it will be referred to as the NP M5-brane theory.
While the Nambu-Poisson bracket is a higher dimensional analogue of the Poisson bracket,
its higher order deformation analogous to the Moyal bracket 
is obstructed by a no-go theorem \cite{Chen:2010ny}.
Via dualities, 
the NP M5-brane theory allows us to understand the effective theory of 
D-branes in large R-R field backgrounds
\cite{Ho,Ho:2013paa,Ma:2012hc}.
In particular, 
to the Poisson level,
the effective Lagrangian for a D$p$-brane in 
the background of a large constant R-R $(p-1)$-form potential
has been derived \cite{Ho:2013paa}.
(Generalizations to non-linear DBI-like theories were proposed 
in Ref.\cite{BP}.)
Setting $p = 3$, 
we have the S-dual of a D3-brane in $B$-field background.
Thus the real challenge is to extend the duality 
for the case of $p = 3$
to higher order corrections beyond the Poisson limit.

More specifically,
D3-branes in both backgrounds can be obtained by compactifying an M5-brane 
in the $C$-field background on a torus.
At the Poisson level, 
the existence of the NP M5-brane theory ensures the equivalence between 
the D3-brane theories in NS-NS and R-R backgrounds,
and both theories are known.
In fact, 
people \cite{Ganor:2000my,Rey:2000hh} have computed the electromagnetic dual theory 
of the noncommutative gauge theory in a perturbative approach, 
using the Seiberg-Witten map \cite{Seiberg:1999vs},
which replaces the noncommutative $U(1)$ symmetry by the Abelian $U(1)$ symmetry.
Our approach differs from theirs in an important way.
The NP M5-brane theory allows us to perform the duality transformation 
without using the Seiberg-Witten map, 
instead we work on the noncommutative fields directly.
As a result, the Lagrangian as an expansion of the coupling constant 
and noncommutativity parameter is quite different.
Incidentally,
there are other works in the literature that dealt with 
a different decoupling limit in which the D3-brane can no longer be described 
by an ordinary field theory \cite{NCOS}.
These works are not directly related to our work presented here,
and our calculations are purely field-theoretical.

In this work, 
we extend the Poisson-level result {\em to all orders}
in the derivative expansion in a compact form.
That is, we have the exact equivalence (S-duality) 
between the D3-brane theories in the NS-NS and R-R field backgrounds. 
The calculation is restricted to the gauge field sector for simplicity.
We are also able to construct the new gauge transformation laws,
as well as the definitions of 
some of the 
covariant field strengths, 
to all orders for the D3-brane in R-R 2-form potential background.

The S-duality of a D3-brane in NS-NS and R-R backgrounds 
was also studied in Ref.\cite{Cornalba:2002cu} from a different perspective.
There, 
infinitesimal NS-NS and R-R backgrounds were turned on simultaneously,
and S-duality was studied within the same theory as a self-dual theory. 
In this work, 
the NS-NS and R-R fields are not turned on at the same time,
and S-duality is studied as a duality between two different theories.

The importance of this result goes beyond the context of string theory.
``What is the electromagnetic dual of noncommutative gauge theory?''
is an old question extensively explored in the literature.
This work demonstrates an example of the electromagnetic duality 
that can be made explicit for a non-Abelian gauge theory.
Hopefully, 
this work will shed some new light 
on the electromagnetic duality for a generic noncommutative Yang-Mills theory.

The plan of this paper is to first review 
the NP M5-brane theory in large $C$-field background
and the D$p$-brane theory in R-R $(p-1)$-form background in Sec. \ref{2}.
A proof of the S-duality at the leading order
(in the Poisson limit) for the D3-brane is given in Sec. \ref{3}.
In Sec. \ref{coupling-inv}, 
we comment on the difference between our approach 
and the perturbative approach using Seiberg-Witten map \cite{Ganor:2000my,Rey:2000hh}.
The calculation is extended to all orders in Sec. \ref{4}.
Our conjectures of the all-order expressions of 
gauge transformation laws and covariant field strengths 
for a D3-brane in R-R 2-form potential are given in Sec. \ref{5}. 
Finally, we conclude in Sec. \ref{6}.

We will restrict ourselves to the cases when 
all time-like components $B_{0i}$ of the $B$-field vanish.
It is then possible to choose spatial coordinates such that
the only non-zero components of $B$ are $B_{\dot1\dot2}$ and $B_{\dot2\dot1}$
for some directions $x^{\dot1}$ and $x^{\dot2}$.


\section{Review of branes in background fields}
\label{2}

In this section we review the effective field theory 
for an M5-brane in large constant $C$-field background, 
and the effective field theory for a D$p$-brane 
in large constant R-R $(p-1)$-form potential background.
These theories are related by dualities, 
and their gauge symmetries are characterized by 
a generalization of the Poisson bracket, 
called Nambu-Poisson bracket, 
and its generalizations.

\subsection{M5-brane in large $C$-field background}

The effective Lagrangian for an M5-brane in large constant $C$-field background 
in flat spacetime \cite{M51,M52,Ho:2009zt}
is given by
\be
{\cal L} = \frac{T_{M5}}{g^2}
\left( {\cal L}_{\mbox{\em \tiny gauge}} + {\cal L}_{\mbox{\em \tiny matter}} \right),
\label{M5S}
\ee
where $T_{M5}$ is the M5-brane tension, 
and
\begin{eqnarray}
{\cal L}_{\mbox{\em \tiny gauge}}
&=&
-\frac{1}{4}{\cal H}_{\a\dot\mu\dot\nu}{\cal H}^{\a\dot\mu\dot\nu}
-\frac{1}{12}{\cal H}_{\dot\mu\dot\nu\dot\lambda}{\cal H}^{\dot\mu\dot\nu\dot\lambda}
\nonumber\\&&
+ \epsilon^{\a\b\g}\epsilon^{\dot\mu\dot\nu\dot\lambda}
\left[ -\frac{1}{2}
\partial_{\dot\mu}b_{\a\dot\nu}\partial_\b b_{\g\dot\lambda}
+\frac{g}{6}
\partial_{\dot\mu}b_{\b\dot\nu}
\epsilon^{\dot\rho\dot\sigma\dot\tau}
\partial_{\dot\sigma}b_{\g\dot\rho}
(\partial_{\dot\lambda}b_{\a\dot\tau}-\partial_{\dot\tau}b_{\a\dot\lambda})
\right].
\label{Sgauge}
\end{eqnarray}
The part of the action related to the matter fields is omitted here 
as we will focus our attention on the gauge field sector in this paper.
The complete M5-brane action (eq.(\ref{M5S})) has
translational symmetry, $SO(2,1)\times SO(3)$ rotation symmetry,
volume-preserving diffemorphism (VPD),
and the 6-dimensional ${\cal N}$ = (2, 0) supersymmetry.

Let us explain our notation.
The anti-symmetric tensor field $b_{AB}$ 
($A, B = 0, 1, \cdots, 5$)
is the gauge potential for the volume-preserving diffeomorphism 
of the 3-dimensional subspace selected by the background $C$-field.
The dotted indices
$\dm, \dn = 3, 4, 5$
denote the directions in which the $C$-field component dominates.\footnote{
We choose the coordinate system such that 
the only non-zero components of the $C$-field are
$C_{345}$ and $C_{012}$ 
with $C_{345} \gg C_{012}$.
}
From time to time, 
we will also use $\dot1, \dot2, \dot3$, 
instead of $3, 4, 5$, 
as the values for the indices $\dm, \dn$.
The rest of the world-volume directions are
labelled by undotted indices $\a, \b = 0, 1, 2$.
The component $C_{012}$
is determined by $C_{345}$ through the nonlinear self-duality relation
on M5-brane.
The signature of spacetime is $\eta=\mbox{diag}(-+\cdots +)$.

The action is a good approximation of the M5-brane
in the large $C$-field limit
defined by the scaling relations
\cite{Chen:2010br}
\be
\ell_P \sim \eps^{1/3}, 
\qquad
g_{(M)\a\b} \sim \eps^{0},
\qquad
g_{(M)\dot\mu\dot\nu} \sim \eps
\qquad
C_{\dot1\dot2\dot3} \sim \eps^0
\label{limit_C}
\ee
with $\eps \rightarrow 0$,
where $\ell_P$ is the Planck length
and $g_{(M)}$ is the spacetime metric in M theory.
The scaling limit (\ref{limit_C})
implies that $C_{012}$ is negligible \cite{Chen:2010br}.

The gauge symmetry of volume-preserving diffeomorphism
is characterized by the Nambu-Poisson bracket $\{\cdot, \cdot, \cdot\}$
defined by
\be
\{f, g, h\} = \eps^{\dm\dn\dlam}\del_{\dm}f \del_{\dn}g \del_{\dlam}h.
\ee
Here $\epsilon_{\dot\mu\dot\nu\dot\lam}$ is the totally anti-symmetrized tensor.

In terms of the variables
\bea
b^{\dm} \equiv \frac{1}{2} \eps^{\dm\dn\dlam} b_{\dn\dlam}, 
\qquad
X^{\dot\mu}(x) \equiv
\frac{x^{\dot\mu}}{g} + b^{\dm},
\label{bXB}
\eea
the covariant field strengths are defined by \cite{M51,M52,Ho:2009zt}
\begin{eqnarray}
{\cal H}_{\a\dot\mu\dot\nu}
&=&\epsilon_{\dot\mu\dot\nu\dot\lambda}{\cal D}_\a X^{\dot\lambda}
\nonumber\\
&=&H_{\a\dot\mu\dot\nu}
-g\epsilon^{\dot\sigma\dot\tau\dot\rho}
(\partial_{\dot\sigma}b_{\a\dot\tau})
\partial_{\dot\rho}b_{\dot\mu\dot\nu},\label{h12def}\\
{\cal H}_{\dot1\dot2\dot3}
&=&g^2\{X^{\dot1},X^{\dot2},X^{\dot3}\}-\frac{1}{g}
\nonumber\\
&=&H_{\dot1\dot2\dot3}
+\frac{g}{2}
(\partial_{\dot\mu}b^{\dot\mu}\partial_{\dot\nu}b^{\dot\nu}
-\partial_{\dot\mu}b^{\dot\nu}\partial_{\dot\nu}b^{\dot\mu})
+g^2\{b^{\dot1},b^{\dot2},b^{\dot3}\},
\label{h30def}
\end{eqnarray}
where $H_{\dot1\dot2\dot3}$ is the classical (undeformed) field strength 
for the Abelian gauge symmetry:
\begin{eqnarray}
H_{\a\dot\mu\dot\nu}
&=&
\partial_{\a}b_{\dot\mu\dot\nu}
-\partial_{\dot\mu}b_{\a\dot\nu}
+\partial_{\dot\nu}b_{\a\dot\mu},\\
H_{\dot\lambda\dot\mu\dot\nu}
&=&
\partial_{\dot\lambda}b_{\dot\mu\dot\nu}
+\partial_{\dot\mu}b_{\dot\nu\dot\lambda}
+\partial_{\dot\nu}b_{\dot\lambda\dot\mu}.
\end{eqnarray}

The gauge transformation laws for the gauge potentials are \cite{M51,M52,Ho:2009zt}
\begin{eqnarray}
\delta_{\Lambda}b_{\dot\mu\dot\nu}
&=&\partial_{\dot\mu}\Lambda_{\dot\nu}
-\partial_{\dot\nu}\Lambda_{\dot\mu}
+g\kappa^{\dot\lambda}\partial_{\dot\lambda} b_{\dot\mu\dot\nu},
\label{gt2}
\\
\delta_{\Lambda} b_{\a\dot\mu}
&=&\partial_{\a}\Lambda_{\dot\mu}
-\partial_{\dot\mu}\Lambda_{\a}
+g\kappa^{\dot\nu}\partial_{\dot\nu}b_{\a\dot\mu}
+g(\partial_{\dot\mu}\kappa^{\dot\nu})b_{\a\dot\nu}, \label{gt4}
\end{eqnarray}
where $\Lambda$ is the gauge parameter and 
\be
\kappa^{\dot\lambda}\equiv
\epsilon^{\dot\lambda\dot\mu\dot\nu}\partial_{\dot\mu}
\Lambda_{\dot\nu}.
\label{def-kappa}
\ee
The first two terms in (\ref{gt2}) and (\ref{gt4}) 
are the classical transformations of a 2-form potential.
In terms of the variables (\ref{bXB}) and 
\be
B_{\a}{}^{\dm} \equiv \eps^{\dm\dn\dlam}\del_{\dn}b_{\a\dlam},
\ee
eqs. (\ref{gt2}) and (\ref{gt4}) can also be rewritten as
\bea
\d_{\Lambda} b^{\dm} &=& \kappa^{\dm} + g\kappa^{\dn}\del_{\dn} b^{\dm},
\\
\d_{\Lambda} B_{\a}{}^{\dm} &=&
\del_{\a}\kappa^{\dm} + g\kappa^{\dn}\del_{\dn}B_{\a}{}^{\dm}
- g(\del_{\dn}\kappa^{\dm})B_{\a}{}^{\dn}.
\eea

\subsection{D$p$-brane in R-R $(p-1)$-form background}

\subsubsection{Gauge symmetry}

Through M-\IIA duality and T-dualities, 
the effective M5-brane theory in large $C$-field background
is related to the effective theory of D$p$-brane 
in the background of a large constant R-R $(p-1)$-form gauge potential \cite{Ho:2013paa}.
Here we are only concerned with the gauge field sector of the theory.

It is well known that the massless mode of an open string ending on a D-brane 
is represented by a massless vector field 
that serves as the gauge potential of the $U(1)$ gauge symmetry.
Similarly,
the massless mode of an open D$(p-2)$-brane ending on the D$p$-brane 
is represented by a massless tensor field.
It serves as the gauge potential of 
the diffeomorphism preserving the constant R-R $(p-1)$-form potential background
\cite{Ho:2013paa}.
Referring to the R-R potential as the volume-form of the $(p-1)$-dimensional subspace,
we identify the gauge symmetry introduced by the open D$(p-2)$-branes
with the volume-preserving diffeomorphism.
This is analogous to how VPD is introduced on a 3-dimensional subspace of 
the M5-brane world-volume by the $C$-field background.

As a generalization of the Nambu-Poisson bracket,
the VPD of a $(p-1)$-dimensional space 
is generated by a $(p-1)$-bracket
\be
\{f_1, f_2, \cdots, f_{p-1}\} \equiv
\eps^{\dm_1 \dm_2 \cdots \dm_{p-1}} 
(\del_{\dm_1}f_1) (\del_{\dm_2}f_2) \cdots (\del_{\dm_{p-1}}f_{p-1}),
\ee
where $\epsilon^{\dm_1 \dm_2 \cdots \dm_{p-1}}$
is the totally anti-symmetrized tensor.
Again we use the dotted indices 
$\dot\mu, \dot\nu = 2, 3, \cdots, p$
for the subspace on which VPD is defined
(i.e. the subspace selected by the background R-R field),
and we will identify $\dot1, \dot2, \dot3, \cdots$ with $2, 3, 4, \cdots$.
The remaining dimensions are labelled by undotted indices 
($\alpha, \beta = 0, 1$).
We will use Latin letters in upper case
($A, B = 0, 1, 2, \cdots, p$)
for all directions of the world-volume.

A field $\Phi$ is {\em VPD-covariant} if it transforms as
\be
\d\Phi = \{f_1, f_2, \cdots, f_{p-2}, \Phi\} = \kappa^{\dm}\del_{\dm}\Phi,
\ee
where the transformation parameter $\kappa^{\dm}$
is defined by a set of $(p-2)$ arbitrary smooth functions $f_1, f_2, \cdots, f_{p-2}$ as
\be
\kappa^{\dm} = 
\eps^{\dm_1\cdots\dm_{p-2}\dm}(\del_{\dm_1}f_1)\cdots(\del_{\dm_{p-2}}f_{p-2})
\ee
and so it is divergenceless
\be
\partial_{\dot\mu}\kappa^{\dot\mu}=0.
\ee

The $(p-2)$-form gauge potential 
$b_{\dm_1\cdots\dm_{p-2}}$ for the gauge symmetry of 
volume-preserving diffeomorphism
is Hodge-dual to a vector field in the $(p-1)$-dimensional subspace \cite{Ho,Ho:2013paa}
\be
b^{\dm_1} = \frac{1}{(p-2)!}\eps^{\dm_1 \dm_2 \cdots \dm_{p-1}} 
b_{\dm_2\cdots\dm_{p-1}}.
\ee
The familiar $U(1)$ gauge symmetry on a D-brane is still present,
although the $U(1)$ field strength has to be modified 
in order for them to be VPD-covariant.
The gauge transformation laws for the VPD gauge potential $b^{\dm}$ 
and the $U(1)$ gauge potential $a_A$ are defined by \cite{Ho,Ho:2013paa}
\bea
\delta_{\Lambda} b^{\dm}&=&
\kappa^{\dm}+g\kappa^{\dn}\del_{\dn}b^{\dm}\label{transf-bdm},\\
\delta_{\Lambda} a_{A}&=&
\del_{A}\lambda+
g(\kappa^{\dn}\del_{\dn}a_{A}+a_{\dn}\del_{A}\kappa^{\dn}),
\label{transf-adm}
\eea
where $(\lambda, \kappa)$ are the $U(1)$ and VPD gauge parameters, 
respectively.
The transformation law for $b$ 
can be obtained by demanding the VPD-covariance of \cite{Ho,Ho:2013paa}
\be
X^{\dm} \equiv \frac{y^{\dm}}{g} + b^{\dm}.
\ee

It is convenient to define the composite field \cite{Ho,Ho:2013paa}
\be
\hat{B}_{\a}{}^{\dm} \equiv
(M^{-1})^{\dm\dn}{}_{\a\b}
(V_{\dn}^{~\ds}\del^{\b}b_{\ds}+\eps^{\b\g}F_{\g\dn}),
\label{def-Bhat}
\ee
where
\bea
V_{\dn}^{~\dm}&\equiv&
g\del_{\dn}X^{\dm} 
= \d_{\dn}^{~\dm}+g\del_{\dn}b^{\dm}, 
\label{def-V} \\
M_{\dm\dn}{}^{\a\b} &\equiv&
V_{\dm\dr}V_{\dn}{}^{\dr}\d^{\a\b}-g\eps^{\a\b}F_{\dm\dn}.
\eea
It is straightforward to compute the gauge transformation law for this composite field:
\be
\delta_{\Lambda} \hat{B}_{\a}^{~\dm} = 
\del_{\a}\kappa^{\dm}+
g(\kappa^{\dn}\del_{\dn}\hat{B}_{\a}^{~\dm}-
\hat{B}_{\a}^{~\dn}\del_{\dn}\kappa^{\dm}).
\label{transf-B}
\ee
It allows us to define VPD-covariant derivatives 
together with $b^{\dm}$:
\bea
{\cal D}_{\alpha}\Phi &\equiv&
\partial_{\alpha}\Phi -g\hat{B}_{\alpha}{}^{\dot\mu}\partial_{\dot\mu}\Phi, \\
{\cal D}_{\dot\mu_1}\Phi &\equiv&
\frac{(-1)^p}{(p-2)!}g^{p-2}\epsilon_{\dot\mu_1\dot\mu_2 \cdots \dot\mu_{p-1}}
\{X^{\dot\mu_2}, X^{\dot\mu_3}, \cdots, X^{\dot\mu_{p-1}}, \Phi\}.
\eea

The VPD-covariant field strength ${\cal H}$ is defined as \cite{Ho,Ho:2013paa}
\be
{\cal H}^{\dm_1\dm_2\cdots\dm_{p-1}} \equiv 
g^{p-2}\{X^{\dm_1}, X^{\dm_2}, \cdots, X^{\dm_{p-1}}\}-\frac{1}{g}\epsilon^{\dm_1\dm_2\cdots\dm_{p-1}}.
\ee
Since the potential $b^{\dm}$ is only defined on the $(p-1)$-dimensional subspace,
the field strength has only one independent component ${\cal H}_{23\cdots p}$.

The VPD-covariant $U(1)$ field strengths are defined by \cite{Ho:2013paa}
\bea
{\cal F}_{\dm\dn}&\equiv&
\frac{g^{p-3}}{(p-3)!} \eps_{\dm\dn\dm_1\cdots\dm_{p-3}}
\{X^{\dm_1}, \cdots, X^{\dm_{p-3}}, a_{\dot{\rho}}, y^{\dot{\rho}} \},
\label{Fdmdn1} \\
{\cal F}_{\a\dm}&\equiv&
{V^{-1}}_{\dm}^{~\dn}(F_{\a\dn}+gF_{\dn\dd}\hat{B}_{\a}^{~\dd}), 
\label{Fadm1} \\
{\cal F}_{\a\b}&\equiv&
F_{\a\b}+g[-F_{\a\dm}\hat{B}_{\b}^{~\dm}-
F_{\dm\b}\hat{B}_{\a}^{~\dm}+
gF_{\dm\dn}\hat{B}_{\a}^{~\dm}\hat{B}_{\b}^{~\dn}],
\label{Fab1}
\eea
where $F_{AB} \equiv \del_A a_B - \del_B a_A$ is usual Abelian field strength,
and $\hat{B}_{\a}{}^{\dm}$ and $V_{\dm}{}^{\dn}$ are defined by
(\ref{def-Bhat}) and (\ref{def-V}), respectively.
For simplicity, 
we consider the theory in configurations where 
all matter fields are turned off.
In particular, 
eq.(\ref{def-Bhat}) should be corrected by terms 
involving matter fields in the complete theory.

In addition to the covariant derivatives and field strengths,
a general class of VPD-covariant, $U(1)$-invariant quantities is given by the formula
\cite{Ho:2013paa}
\be
{\cal O}_{nml} \equiv
\{X^{\dm_1}, \cdots, X^{\dm_n}, 
a_{\dn_1}, \cdots, a_{\dn_m}, \frac{y^{\dn_1}}{g}, \cdots, \frac{y^{\dn_m}}{g}, \Phi_1, \cdots, \Phi_l\},
\label{Onml} 
\ee
where $n, m, l \geq 0$ and $n + 2m + l = p-1$,
for an arbitrary set of covariant quantities $\Phi_I$ $(I = 1, 2, \cdots, l)$.
In fact, 
${\cal D}_{\dm}\Phi$, ${\cal H}_{2\cdots p}$ and ${\cal F}_{\dm\dn}$
all belong to this class.

\subsubsection{Lagrangian}

The effective Lagrangian of a D$p$-brane
in large constant R-R $(p-1)$-form background is \cite{Ho:2013paa}
\be
{\cal L} = T_p ({\cal L}_{\mbox{\em \tiny gauge}} + {\cal L}_{\mbox{\em \tiny matter}}),
\label{L}
\ee
where $T_p$ is an overall constant and
\bea
{\cal L}_{\mbox{\em \tiny gauge}} 
&\equiv& 
- \frac{1}{2(p-1)!}
\left({\cal H}_{\dm_1\cdots\dm_{p-1}} + \frac{1}{g}\eps_{\dm_1\cdots\dm_{p-1}}\right)^2
+ \frac{1}{2g}\eps^{\a\b}{\cal F}_{\a\b}
+ \frac{1}{2}{\cal F}_{\a\dm}^2
- \frac{1}{4} {\cal F}_{\dm\dn}^2
\nn \\
&&
- \frac{1}{2} \sum_{m=2}^{\lfloor \frac{p-1}{2} \rfloor}
\frac{g^{2(p-m-2)}}{((p-2m-1)!)(m!)^2} 
\{ X^{\dm_1}, \cdots, X^{\dm_{p-2m-1}}, a_{\dn_1}, \cdots, a_{\dn_m}, y^{\dn_1}, \cdots, y^{\dn_m}\}^2.
\nn \\
\label{L-Dp}
\eea
We omit the Lagrangian for the matter fields.

The first term in ${\cal L}_{\mbox{\em \tiny gauge}}$ 
is the kinetic term for the field strength ${\cal H}$ 
with a large background field of order $1/g$ turned on.
The second is the Wess-Zumino interaction term
describing the coupling of the field strength ${\cal F}_{01}$
to the R-R field background $C_{23\cdots p} = 1/g$.
The third and fourth terms in ${\cal L}_{\mbox{\em \tiny gauge}}$ 
look like part of the standard kinetic term 
for the $U(1)$ gauge field ${\cal F}_{AB}$, 
but there is a relative minus sign, 
with a kinetic term
$\frac{1}{4}{\cal F}_{\a\b}^2$
missing.

To see how the correct kinetic term is hidden in the formula,
we examine the perturbative expansion of the Lagrangian in powers of $g$.
In particular, 
the Wess-Zumino term is
\bea
\frac{1}{2g}\eps^{\a\b}{\cal F}_{\a\b} 
&\simeq& 
- F_{01}H_{23\cdots p} - F_{\a\dm}^2 + \cdots,
\label{WZ-g}
\eea
where we ignored total derivatives,
and the Abelian field strengths $H$, $F$ 
are the 0-th order approximations of ${\cal H}$ and ${\cal F}$.
Note that the second term above flips the sign 
of the third term in ${\cal L}_{\mbox{\em \tiny gauge}}$ at the 0-th order.
To the 0-th order in $g$, 
${\cal L}_{\mbox{\em \tiny gauge}}$ is hence
\be
{\cal L}_{\mbox{\em \tiny gauge}} =
-\frac{1}{2}(H_{23\cdots p}+F_{01})^2
-\frac{1}{4}F_{AB}F^{AB}
+ \mbox{total derivatives} + {\cal O}(g).
\label{Sgauge10Dp}
\ee
Since the time-derivative of $H_{23\cdots p}$ does not appear in the Lagrangian,
we can integrate it out in the perturbation theory, 
and obtain the Maxwell theory in $(p+1)$ dimensions at the lowest order.
Therefore, 
while the gauge potential $b^{\dm}$ is present in the theory to represent 
the massless mode of an open D$(p-2)$-brane ending on D$p$-brane,
it is dual to (part of) the $U(1)$ gauge potential $a_A$ 
without introducing additional physical degrees of freedom.

The scaling limit for the effective Lagrangian given above to provide 
a good approximation of a D$p$-brane in R-R field background is \cite{Ho:2013paa}
\bea
&\ell_s \sim \eps^{1/2},
\qquad
g_s \sim \eps^{-1/2},
\qquad
C_{\dm_1\cdots\dm_{p-1}} \sim 1,
\\
&g_{\a\b} \sim 1, \quad (\a, \b = 0, 1)
\qquad
g_{\dm\dn} \sim \eps, \quad (\dm, \dn = 2, 3, \cdots, p)
\eea
with $\eps \rightarrow 0$.


\section{From NS-NS to R-R}
\label{3}

Starting with the NP M5-brane theory,
one can perform double dimensional reduction (DDR) twice along two different directions 
to get a D3-brane in constant NS-NS or R-R background,
depending on the ordering of dimensional reductions \cite{M52, Ho}.
For a constant $C$-field background 
with its dominant component lying
in the directions of $x^{\dot1}, x^{\dot2}, x^{\dot3}$
(also denoted as $x^3, x^4, x^5$,
with the rest of the world-volume coordinates denoted by $x^0, x^1, x^2$)
one can first do a double dimensional reduction along $x^{\dot3}$ 
to obtain a D4-brane in the NS-NS $B$-field background.
Then another dimensional reduction along $x^2$ leads to 
a D3-brane in the NS-NS $B$-field background via T-duality. 
On the other hand, 
if we switch the ordering of the two dimensional reductions, 
we can first get a D4-brane in R-R $C$-field background via DDR along $x^2$,
and then obtain a D3-brane in R-R two-form background via DDR along $x^{\dot3}$. 
Since the ordering of dimensional reductions should not change anything physical, 
the resulting theories must be equivalent.
This is the argument for the S-duality of D3-brane theory.

In this section we show that, in the Poisson limit,
the D3-brane theory in large NS-NS $B$-field background 
is equivalent to that in large R-R two-form potential background.
This is conceptually guaranteed by the existence of the NP M5-theory.
We will extend the result beyond the Poisson limit, 
to all orders in the large background expansion, 
in the next section.

The gauge field part of the world-volume Lagrangian 
for a D3-brane in the background of 
a large NS-NS two-from potential in the Poisson limit is
given by that of a noncommutative $U(1)$ gauge theory
\cite{Chu-Ho,Schomerus:1999ug,Seiberg:1999vs}:
\bea
{\cal L}_{NS-NS} \equiv
-\frac{1}{4}{\cal F}_{\alpha\beta}^{\prime}{\cal F}^{\prime\alpha\beta}
-\frac{1}{2}{\cal F}_{\alpha\dot\mu}^{\prime}{\cal F}^{\prime\alpha\dot\mu}
-\frac{1}{4}{\cal F}_{\dot\mu\dot\nu}^{\prime}{\cal F}^{\prime\dot\mu\dot\nu}.
\label{action-NSNS}
\eea
(The overall constant factor is omitted.)
We refer to this theory as the NS-NS theory, 
and correspondingly there is an R-R theory.
To distinguish quantities in these two theories, 
we apply primes to quantities in the NS-NS theory,
and use symbols without primes for quantities in the R-R theory.
Our convention of world-volume indices is that 
$\a, \b = 0, 1$, $\dot\mu, \dot\nu = \dot1, \dot2$ 
and $A, B = 0, 1, \dot1, \dot2$. 
The dotted indices label the directions in which 
the NS-NS $B$-field (or R-R field) background is turned on.
(We assume that all time-like components of the $B$-field vanish.)
The Poisson limit of the noncommutative gauge field strengths is given by
\bea
{\cal F}^{\prime}_{AB} 
\equiv
F^{\prime}_{AB}+g\{a^{\prime}_A, a^{\prime}_B\},
\eea
where $F^{\prime}_{AB} \equiv \partial_A a^{\prime}_B-\partial_B a^{\prime}_A$
and the Poisson bracket is defined by
\be
\{f_1(x), f_2(x)\} \equiv 
\epsilon^{\dot\mu\dot\nu} \del_{\dot\mu}f_1 \del_{\dot\nu}f_2,
\ee
where $\eps^{\dot\mu\dot\nu}$ is the totally anti-symmetrized tensor.

For small $g$, 
both theories dual to each other can be expanded in powers of $g$.
There is an overall factor of the Lagrangian that has been omitted.
It is straightforward to re-insert the coupling constant everywhere 
in our derivation.
We will ignore it in this section for simplicity.

The Lagrangian in the Poisson limit for the R-R theory is known \cite{Ho,Ho:2013paa}, 
with the gauge field part given by
\bea
\label{action-RR}
{\cal L}_{RR} \equiv 
-\frac{1}{2}{\cal H}^2_{\dot1\dot2}+\frac{1}{2}{\cal F}_{\alpha\dot\mu}{\cal F}^{\alpha\dot\mu}
-\frac{1}{4}F_{\dot\mu\dot\nu}F^{\dot\mu\dot\nu}
+\frac{1}{2g}\epsilon^{\alpha\beta}{\cal F}_{\alpha\beta}
\eea
up to total derivatives.
Once again we ignore the overall constant factor in the Lagrangian.
The field strengths are defined as
\bea
{\cal H}_{\dot1\dot2}&\equiv&
H_{\dot1\dot2}+g\{b_{\dot1}, b_{\dot2}\},
\label{H-P} \\
{\cal F}_{\alpha\dot\mu}&\equiv&
(V^{-1})_{\dot\mu}{}^{\dot\nu}\bigg(F_{\alpha\dot\nu}
+gF_{\dot\nu\dot\sigma}\hat{B}_{\alpha}{}^{\dot\sigma}\bigg),
\label{Fadm-P} \\
{\cal F}_{\alpha\beta}&\equiv&
F_{\alpha\beta}+
g\bigg(
-F_{\alpha\dot\mu}\hat{B}_{\beta}{}^{\dot\mu}
-F_{\dot\mu\beta}\hat{B}_{\alpha}{}^{\dot\mu}
+gF_{\dot\mu\dot\nu}\hat{B}_{\alpha}{}^{\dot\mu}\hat{B}_{\beta}{}^{\dot\nu}
\bigg),
\label{Fab-P}
\eea
where
\bea
H_{\dot1\dot2} &\equiv&
\del_{\dot\mu}b^{\dot\mu},
\\
V_{\dot\mu}{}^{\dot\nu}&\equiv&
\delta_{\dot\mu}{}^{\dot\nu}+g\eps^{\dot\nu\dot\lam}\partial_{\dot\mu}b_{\lam},
\\
F_{AB}&\equiv&
\partial_A a_B-\partial_B a_A,
\eea
and $\hat{B}_{\alpha}{}^{\dot\mu}$ is defined as the solution to the equation
\bea
V_{\dot\mu}{}^{\dot\nu}
\bigg(\partial^{\alpha}b_{\dot\nu}-V^{\dot\rho}{}_{\dot\nu}\hat{B}^{\alpha}{}_{\dot\rho}\bigg)
+\epsilon^{\alpha\beta}F_{\beta\dot\mu}+g\epsilon^{\alpha\beta}F_{\dot\mu\dot\nu}
\hat{B}_{\beta}{}^{\dot\nu}=0.
\label{Bhat}
\eea
Here and below we use the Lorentzian metric $\eta^{AB}$ 
to raise or lower indices.
For example, $V^{\dot\rho}{}_{\dot\nu}$ is defined as
$\eta^{\dot\rho\dot\lam}\eta_{\dot\nu\dot\mu}V_{\dot\lam}{}^{\dot\mu}$,
which equals $V_{\dot\rho}{}^{\dot\nu}$ 
since $\eta^{\dot\mu\dot\nu} = \delta^{\dot\mu\dot\nu}$.
Strictly speaking,
the relation (\ref{Bhat}) holds only when the matter fields vanish on the D3-brane.
We have consistently ignored all matter field contributions to the theory, 
and leave the complete theory for future works.

To show the S-duality between the NS-NS theory and the R-R theory 
in the Poisson limit,
let us carry out a step-by-step proof of the duality.
As the first step, 
the gauge field variable $b^{\dot\mu}$ in the R-R theory
can be identified with $a'_{\dot\mu}$ in the NS-NS theory via
\be
b^{\dot\mu}\equiv\epsilon^{\dot\mu\dot\nu}a_{\dot\nu}^{\prime}.
\label{change-b-a}
\ee
As a result,
\bea
{\cal F}_{\dot1\dot2}^{\prime}
=F_{\dot1\dot2}^{\prime}+g\{ a_{\dot1}^{\prime}, a_{\dot2}^{\prime}\}
={\cal H}_{\dot1\dot2}.
\label{identify-F-H}
\eea
One can also check that
\bea
\label{B1}
{\cal F}^{\prime}_{\a\dot\mu}=
- \epsilon_{\dot\mu\dot\nu}\left(
\partial_{\alpha}b_{\dot\nu} 
- V_{\dot\lambda}{}^{\dot\nu}B_{\alpha}{}^{\dot\lambda}
\right),
\eea
where
\be
B_{\alpha}{}^{\dot\mu}\equiv
\epsilon^{\dot\mu\dot\nu}\partial_{\dot\nu}a'_{\alpha}.
\label{def-B}
\ee
Note that the notation here is different from Ref.\cite{Ho} 
as we use the metric to raise and lower indices for all quantities, 
including $b_{\dot\mu}$.

Let us also define
\bea
\label{B2}
{\cal F}^{\prime\prime}_{\alpha\dot\mu}\equiv
\epsilon_{\alpha\beta}{\cal F}^{\prime\beta\dot\nu}\epsilon_{\dot\nu\dot\mu}
= \epsilon_{\alpha\beta}\left(
\partial^{\beta}b_{\dot\mu} - B^{\beta\dot\nu}V_{\dot\nu\dot\mu}
\right),
\eea
which will be identified with ${\cal F}_{\alpha\dot\mu}$
in the R-R theory later after certain duality transformations.
After the change of variables, 
the Lagrangian (\ref{action-NSNS}) is equivalent to
\bea
\label{action3}
{\cal L}_{NS-NS} = 
-\frac{1}{4}{\cal F}_{\alpha\beta}^{\prime}{\cal F}^{\prime\alpha\beta}
+\frac{1}{2}{\cal F}^{\prime\prime}_{\alpha\dot\mu}{\cal F}^{\prime\prime\alpha\dot\mu}
-\frac{1}{2}{\cal H}_{\dot1\dot2}{\cal H}^{\dot1\dot2}.
\label{action-1}
\eea

The next step is to dualize the field component ${\cal F}_{\a\b}$.
The Lagrangian above is equivalent to the following Lagrangian
\bea
\label{action1}
{\cal L}_{NS-NS}^{(1)} \equiv
-\frac{1}{2}\phi^2+\frac{1}{2}\epsilon^{\a\b}{\cal F}_{\a\b}^{\prime}\phi
+\frac{1}{2}{\cal F}^{\prime\prime}_{\alpha\dot\mu}{\cal F}^{\prime\prime\alpha\dot\mu}
-\frac{1}{2}{\cal H}_{\dot1\dot2}{\cal H}^{\dot1\dot2}.
\eea
The reason is that the equation of motion of $\phi$ imposes the constraint 
\be
\phi = {\cal F}'_{01}.
\ee
Replacing $\phi$ by ${\cal F}'_{01}$ in this Lagrangian 
reproduces (\ref{action-1}).

It is less obvious that we can use the field strength 
\be
F_{\dot\mu\dot\nu} \equiv \del_{\dot\mu}a_{\dot\nu} - \del_{\dot\mu}a_{\dot\nu},
\ee
in the place of $\phi$.
Yet we claim that the Lagrangian 
\bea
\label{action4}
{\cal L}^{(2)}_{NS-NS} = 
-\frac{1}{2}F_{\dot1\dot2}^2
+\frac{1}{2}\epsilon^{\a\b}{\cal F}_{\a\b}^{\prime}F_{\dot1\dot2}
+\frac{1}{2}{\cal F}^{\prime\prime}_{\alpha\dot\mu}{\cal F}^{\prime\prime\alpha\dot\mu}
-\frac{1}{2}{\cal H}_{\dot1\dot2}{\cal H}^{\dot1\dot2}
\eea
is also equivalent to (\ref{action-1}).
Its main difference from (\ref{action1}) is that 
the equation of motion of the new field $a_{\dot\mu}$ does not imply
\be
F_{\dot1\dot2}={\cal F}_{01}^{\prime},
\label{eom-a1}
\ee
which would allow one to reproduce (\ref{action-1}) from (\ref{action4}),
but only
\be
\partial_{\dot\mu}\bigg(F_{\dot1\dot2}-{\cal F}_{01}^{\prime}\bigg)=0.
\label{eom-a}
\ee 
However, since both $x^{\dot1}$ and $x^{\dot2}$ are spatial coordinates,
suitable boundary conditions at the infinities of the coordinates $x^{\dm}$
would allow us to deduce (\ref{eom-a1}) from (\ref{eom-a}).
Thus we have carried out the duality transformation 
from the field ${\cal F}'_{\alpha\beta}$ to its dual field $F_{\dot1\dot2}$ in the R-R theory.
The consistency of this step relies on the fact that 
there is no nontrivial Bianchi identity for the field strength $F_{\dot1\dot2}$
in the 2-dimensional subspace of $(x^{\dot1}, x^{\dot2})$.

The last step is to carry out the electromagnetic duality transformation 
to get $a_{\alpha}$ from $a'_{\a}$. 
Before that, we expand the 2nd term in the Lagrangian (\ref{action4}) as
\bea
{\cal F}_{01}^{\prime}F_{\dot1\dot2}
&=&
F_{01}^{\prime}F_{\dot1\dot2} + \{a_0^{\prime}, a_1^{\prime}\}F_{\dot1\dot2}
\nn \\
&=& \epsilon^{\alpha\beta}\partial_{\beta}a_{\dot\mu}B_{\alpha}{}^{\dot\mu}
+ \epsilon^{\alpha\beta}F_{\dot1\dot2}B_{\alpha}{}^{\dot1}B_{\beta}^{\dot2}
+ \mbox{total derivatives},
\label{F01F12}
\eea
and ignore the total derivatives.
Then we replace $B_{\alpha}{}^{\dot\mu}$ by $\breve{B}_{\alpha}{}^{\dot\mu}$ 
in the Lagrangian and add an additional term
\bea
\label{additional term}
\epsilon^{\alpha\beta}f_{\beta\dot\mu}
\lbrack \breve{B}_{\alpha}{}^{\dot\mu}
-\epsilon^{\dot\mu\dot\nu}\partial_{\dot\nu}a'_{\alpha}
\rbrack.
\eea
The new Lagrangian is equivalent to the previous one (\ref{action4}) because
the new variable $\breve{B}_{\alpha}{}^{\dot\mu}$ is forced 
to be equal to $B_{\alpha}{}^{\dot\mu}$ (\ref{def-B})
when the Lagrange multiplier $f_{\beta\dot\mu}$ is integrated out.
On the other hand, if we integrate out $a'_{\alpha}$, 
we get $\epsilon^{\dot\mu\dot\nu}\partial_{\dot\nu}f_{\beta\dot\mu}=0$. 
It implies that locally 
$f_{\beta\dot\mu} = - \partial_{\dot\mu}a_{\beta}$ 
for some field $a_{\beta}$. 
As a result, eq.(\ref{additional term}) becomes
\bea
-\epsilon^{\alpha\beta}\partial_{\dot\mu}a_{\beta}\breve{B}_{\alpha}{}^{\dot\mu}.
\eea
One can now easily integrate out $\breve{B}_{\alpha}{}^{\dot\mu}$.
The result is equivalent to replacing $\breve{B}_{\alpha}{}^{\dot\mu}$ by 
the solution of its equation of motion,
which is exactly the same as $\hat{B}_{\alpha}{}^{\dot\mu}$ defined by (\ref{Bhat}).

A difference between $B_{\alpha}{}^{\dot\mu}$ and $\hat{B}_{\alpha}{}^{\dot\mu}$ is 
that the former satisfies the off-shell constraint $\partial_{\dot\mu}B_{\alpha}{}^{\dot\mu} = 0$,
while the latter is divergenceless only when its equation of motion is imposed.
The integrability condition is interchanged with the equation of motion.
This is a typical feature of electromagnetic duality.

It is easy to check that,
after integrating out $\breve{B}_{\alpha}{}^{\dot\mu}$,
eq.(\ref{F01F12}) can be rewritten as
\be
\frac{1}{2g}\epsilon^{\alpha\beta}{\cal F}_{\alpha\beta}
\ee
up to total derivatives.
According to the definition of ${\cal F}^{\prime\prime}_{\alpha\dot\mu}$ (\ref{B2}), 
we also find
\be
{\cal F}^{\prime\prime}_{\alpha\dot\mu}={\cal F}_{\alpha\dot\mu}.
\ee
The Lagrangian (\ref{action4}) is therefore equivalent to
that of the R-R theory (\ref{action-RR})
\bea
{\cal L}_{RR} = -\frac{1}{2}{\cal H}^2_{\dot1\dot2}
+\frac{1}{2}{\cal F}_{\alpha\dot\mu}{\cal F}^{\alpha\dot\mu}
-\frac{1}{4}F_{\dot\mu\dot\nu}F^{\dot\mu\dot\nu}
+\frac{1}{2g}\epsilon^{\alpha\beta}{\cal F}_{\alpha\beta}.
\eea
Hence, we have shown the S-duality at the Poisson level 
for a D3-brane in R-R and NS-NS backgrounds.

\section{Comment on coupling constant}
\label{coupling-inv}

In Ref.\cite{Ganor:2000my},
the electromagnetic duality of a noncommutative $U(1)$ gauge theory 
was carried out perturbatively to the lowest order 
using the Seiberg-Witten map \cite{Seiberg:1999vs}, 
which allows one to replace the noncommutative $U(1)$ gauge symmetry 
by the commutative $U(1)$ gauge symmetry, 
but with a more complicated Lagrangian.
The electromagnetic duality is taken with respect to 
the commutative $U(1)$ field strength, 
of which the original and dual Lagrangians are both infinite expansions.

To compare our result with that of Ref.\cite{Ganor:2000my},
we replace the noncommutativity parameter $g$ by the symbol $\theta'$,
and put back the overall constant in the Lagrangians.
The Lagrangian of the NS-NS theory (\ref{action-NSNS}) is 
\bea
{\cal L}_{NS-NS} \equiv \frac{1}{g'_{G}{}^2} \Big[
-\frac{1}{4}{\cal F}_{\alpha\beta}^{\prime}{\cal F}^{\prime\alpha\beta}
-\frac{1}{2}{\cal F}_{\alpha\dot\mu}^{\prime}{\cal F}^{\prime\alpha\dot\mu}
-\frac{1}{4}{\cal F}_{\dot\mu\dot\nu}^{\prime}{\cal F}^{\prime\dot\mu\dot\nu}
\Big],
\label{action-NSNS-new-notation}
\eea
where $g'_G$ is the gauge coupling and 
the field strength is defined as
\be
{\cal F}'_{AB} = F'_{AB} + \theta' \{a'_A, a'_B\}.
\ee
Our calculation above for the electromagnetic duality 
can be straightforwardly repeated. 
The only change is the insertion of the overall constant $(g'_G)^{-2}$.
As a result,
the Lagrangian of the R-R theory is the same as (\ref{action-RR})
multiplied by the overall constant $(g'_G)^{-2}$.

On the other hand, if we rescale the potentials 
via the replacement
\bea
b^{\dm} \rightarrow g'_G{}^2 b^{\dm}, 
\qquad
B_{\alpha}{}^{\dm} \rightarrow g'_G{}^2 B_{\alpha}{}^{\dm},
\qquad
a_{\dm} \rightarrow g'_G{}^2 a_{\dm},
\qquad
a_{\a} \rightarrow g'_G{}^2 a_{\a},
\eea
we arrive at the same form of the Lagrangian for the R-R theory (\ref{action-RR})
\be
{\cal L}_{RR} \equiv \frac{1}{g_G^2} \Big[
-\frac{1}{2}{\cal H}^2_{\dot1\dot2}+\frac{1}{2}{\cal F}_{\alpha\dot\mu}{\cal F}^{\alpha\dot\mu}
-\frac{1}{4}F_{\dot\mu\dot\nu}F^{\dot\mu\dot\nu}
+\frac{1}{2\theta}\epsilon^{\alpha\beta}{\cal F}_{\alpha\beta}
\Big],
\label{action-RR-new-notation}
\ee
but with the gauge coupling $g_G$ and noncommutativity parameter $\theta$ defined by
\bea
g_G \equiv \frac{1}{g'_G},
\label{gG} \\
\theta \equiv g'_G{}^2 \theta',
\label{theta}
\eea
and with all the coupling constants $g$ replaced by $\theta$ 
in the definition of the field strengths (\ref{H-P}) -- (\ref{Fab-P})
in the R-R theory.

The inversion of the coupling in (\ref{gG}), 
a familiar feature of electromagnetic duality, 
and the change of the noncommutativity parameter from $\theta'$ to $\theta$ in (\ref{theta})
are in agreement with Ref.\cite{Ganor:2000my}.
In Ref.\cite{Ganor:2000my},
Seiberg-Witten map leads to higher order terms in the Lagrangians,
for which $\theta'$ and $\theta$ are the expansion parameters 
of the original and dual theories, respectively.

Notice that the need of two independent parameters 
$(g'_G, \theta')$ or $(g_G, \theta)$ 
can be seen only if higher order terms are included.
At the lowest order, 
as we mentioned above,
one can scale the gauge fields so that 
the R-R action (\ref{action-RR-new-notation}) has 
\be
g_G = g'_G
\qquad \mbox{and} \qquad
\theta = \theta'.
\label{parameters}
\ee

It may seem strange that the gauge coupling is not inverted 
by the electromagnetic duality.
The same situation occurs for the electromagnetic duality 
of the Abelian $U(1)$ gauge theory.
The inversion of the gauge coupling cannot be seen 
directly from the kinetic term of the gauge fields, 
as it can be set to any value by scaling the fields.
Similarly, in our approach, 
the gauge couplings $g_G$ and $g'_G$ are determined by 
the normalization of the gauge fields.
It is possible to normalize the fields such that 
the expansion parameters of both NS-NS and R-R theories are the same.

The different interpretation of the parameters $\theta$ and $\theta'$ 
and the difference in the definition of gauge symmetry in the R-R theory 
distinguish our approach from that of Ref.\cite{Ganor:2000my}.
In our approach, 
the noncommutativity parameter $\theta'$ in the NS-NS theory 
is dualized to the R-R theory parameter $\theta$, 
which no longer characterizes the noncommutativity of the spatial coordinates,
but rather the deformation of the area-preserving diffeomorphism.
It is the gauge symmetry of the deformed area-preserving diffeomorphism, 
instead of the Seiberg-Witten map, 
that constrains how higher order terms are to be introduced in the R-R Lagrangian.
What we will see in the next section is that 
one can write down the all-order expressions of the NS-NS and R-R Lagrangians, 
for which the expansion parameters are the same (denoted by $g$).
In other words, 
the relation (\ref{parameters}) persists naturally to all orders, 
while (\ref{gG}) and (\ref{theta}) are not useful at higher orders.

In the context of string theory,
the NS-NS theory is a good approximation of the D3-brane world-volume theory
in the double scaling limit of Seiberg-Witten \cite{Seiberg:1999vs}:
\be
\ell_s \sim \epsilon^{1/4}, 
\qquad
g_s \sim \epsilon^{1/2},
\qquad
B_{\dm\dn} \sim 1,
\qquad
g_{\a\b} \sim 1, 
\qquad
g_{\dm\dn} \sim \epsilon.
\ee
The R-R theory is a good approximation of the D3-brane world-volume theory 
in the limit (\ref{limit_C}), 
which is precisely the S-dual of the Seiberg-Witten limit.
Despite the fact that the string coupling $g_s$ is large for the R-R theory 
when it is small for the NS-NS theory,
the decoupling of the D3-brane from the bulk in the Seiberg-Witten limit 
ensures that its S-dual picture also has an effective world-volume theory 
decoupled from the bulk.


\section{S-duality to all orders}
\label{4}

In the last section, 
we have re-derived the D3-brane effective action in large R-R field background 
by carrying out a sequence of electromagnetic dualities 
and field redefinitions 
from the gauge theory for D3-brane in NS-NS background, 
but only at the leading order.
In this section, 
we extend the calculations to all orders,
to derive the S-dual Lagrangian for a D3-brane in R-R field background.

Once again we start with the noncommutative gauge theory 
for a D3-brane in NS-NS background (\ref{action-NSNS}),
but now with the field strengths defined by the noncommutative product
\be
{\cal F}'_{AB} \equiv
F'_{AB} + [a'_A, a'_B]_{\ast}.
\ee
Here the commutator responsible for the non-Abelian nature of the gauge theory
\be
[A, B]_{\ast} \equiv A\ast B - B\ast A
\ee
is defined by the Moyal product
\bea
A\ast B &\equiv& 
\hat{m} \Big(e^{(g/2)\epsilon^{\dot\mu\dot\nu}\partial_{\dot\mu}\otimes
\partial_{\dot\nu}}\Big) (A\otimes B)
\nn 
\\
&\equiv&A\exp\Big({\frac{g\epsilon^{\dot\mu\dot\nu}\overleftarrow{\partial}_{\dot\mu}
\overrightarrow{\partial_{\dot\nu}}}{2}}\Big)B,
\eea
where the map $\hat{m}$ is defined to be the classical (commutative) product
\be
\hat{m} (A\otimes B) = AB.
\ee
The same change of variable (\ref{change-b-a}) can be used to identify
${\cal F}'_{\dot1\dot2}$ with 
the all-order expression of the field strength ${\cal H}_{\dot1\dot2}$
\be
{\cal H}_{\dot1\dot2} \equiv
H_{\dot1\dot2} + [b_{\dot1}, b_{\dot2}]_{\ast}
\ee
in the dual theory.
We will study the gauge transformation 
of the dual theory later.
The identity
\be
{\cal F}_{\dot1\dot2}^{\prime}={\cal H}_{\dot1\dot2}
\ee
is preserved to all orders.

We also define $B_{\alpha}{}^{\dot\mu}$ by the same expression (\ref{def-B}).
In terms of the variables $b_{\dot\mu}$ and $B_{\alpha}{}^{\dot\mu}$, 
the expression for ${\cal F}''_{\alpha\dot\mu}$ is now different from (\ref{B2}).
It is
\be
{\cal F}''_{\alpha\dot\mu} \equiv
- \epsilon_{\alpha\beta} [D'{}^{\beta}, X_{\dot\mu}]_{\ast}
=
- \epsilon_{\alpha\beta} \left(
\partial^{\beta} b_{\dot\mu} - g\{ \partial_{\dot\rho}X_{\dot\mu}, B^{\beta\dot\rho} \}_{**} 
\right),
\ee
where
\bea
D'_{\alpha} &\equiv& \partial_{\alpha} + a'_{\alpha}, 
\\
X^{\dot\mu} &\equiv& \frac{x^{\dot\mu}}{g} + b^{\dot\mu}.
\eea
The new bracket introduced above is defined by
\be
\{A, B\}_{**} \equiv
\hat{m} \Big(
\frac{ e^{(g/2)\epsilon^{\dot\mu\dot\nu}\partial_{\dot\mu}\otimes\partial_{\dot\nu}}
- e^{-(g/2)\epsilon^{\dot\mu\dot\nu}\partial_{\dot\mu}\otimes\partial_{\dot\nu}} }
{g\epsilon^{\dot\lambda\dot\rho}\partial_{\dot\lambda}\otimes\partial_{\dot\rho}} \Big)
(A\otimes B).
\ee
This can also be defined in terms of a commutative (but non-associative) product
\be
(A**B) \equiv 
A\frac{\exp({\frac{g\epsilon^{\dot\mu_1\dot\nu_1}\overleftarrow{\partial}_{\dot\mu_1}
\overrightarrow
{\partial_{\dot\nu_1}}}{2}})-1}{g\epsilon^{\dot\mu_2\dot\nu_2}
\overleftarrow{\partial}_{\dot\mu_2}\overrightarrow{\partial}_{\dot\nu_2}}B
\label{non-assoc-product}
\ee
as
\be
\{A,B\}_{**} \equiv (A**B)+(B**A).
\ee
This product ** has appeared in the literature 
\cite{Garousi:1999ch, Liu:2000ad, Liu:2001qa}.

The next step is to identify 
${\cal F}'_{01}$ with $F_{\dot1\dot2}$ in the dual theory.
Here we still define $F_{\dot1\dot2}$ by the classical expression 
for the undeformed $U(1)$ gauge symmetry.
Up to this point, 
we have shown that formally the same action (\ref{action4}) 
is equivalent to the noncommutative gauge theory, 
but with the field strengths defined with higher order corrections.

The all-order analogue of (\ref{F01F12}) is 
\bea
{\cal F}_{01}^{\prime}F_{\dot1\dot2}
&=&
F_{01}^{\prime}F_{\dot1\dot2} + [a_0^{\prime}, a_1^{\prime}]_* F_{\dot1\dot2}
\nn \\
&=& \epsilon^{\alpha\beta}\partial_{\beta}a_{\dot\mu}B_{\alpha}{}^{\dot\mu}
+ g\epsilon^{\alpha\beta}F_{\dot1\dot2} \{B_{\alpha}{}^{\dot1}, B_{\beta}^{\dot2}\}_{**}
+ \mbox{total derivatives}.
\eea
The same terms (\ref{additional term}) should be added to the Lagrangian
to promote $B_{\alpha}{}^{\dot\mu}$ to an independent field $\breve{B}_{\alpha}{}^{\dot\mu}$, 
so that we can dualize $a'_{\alpha}$ to the potential $a_{\alpha}$ in the dual theory.
The latter arises from solving the equation of motion of $f_{\alpha\dot\mu}$,
in the same way we did in the Poisson limit.

We do not deform the definitions related to the field $B_{\alpha}{}^{\dot\mu}$
because if we had introduced the noncommutative algebra into its definition, 
the term (\ref{additional term}) would have been corrected to include 
additional terms involving gauge potentials.
This would have prohibited us to solve $f_{\alpha\dot\mu}$ explicitly from 
its equation of motion, 
and the appearance of the dual potential $a_{\alpha}$ would have been a problem.

Having replaced $f_{\alpha\dot\mu}$ by the solution to its equation of motion,
\be
f_{\alpha\dot\mu} = - \partial_{\dot\mu} a_{\alpha},
\ee
one can also integrate out $\breve{B}_{\alpha}{}^{\dot\mu}$.
Despite the infinitely many derivatives in the all-order expression,
the equation of motion of $\breve{B}_{\alpha}{}^{\dot\mu}$
does not involve time derivatives.
It is hence still legitimate to replace $\breve{B}_{\alpha}{}^{\dot\mu}$
by the solution to its equation of motion.
We will use $\hat{B}$ to denote the solution.
The difference between keeping $\breve{B}$ as a free variable 
and replacing it by $\hat{B}$ 
lies in whether one has solved the equation of motion of $\breve{B}$.

Finally, 
the last step is to define ${\cal F}_{\alpha\beta}$ 
so that the Lagrangian of the dual theory can be written in the form (\ref{action-RR}),
with all field strengths defined with higher order corrections.
Since ${\cal F}_{\alpha\beta}$ appears linearly in the action,
it is determined by the action only up to total derivatives:
\be
{\cal F}_{\alpha\beta} \equiv
F_{\alpha\beta}
+g(-F_{\alpha\dot\mu}\hat{B}_{\beta}{}^{\dot\mu}
+F_{\beta\dot\mu}\hat{B}_{\alpha}{}^{\dot\mu})
+g^2 F_{\dot\mu\dot\nu}
\{\hat{B}_{\alpha}{}^{\dot\mu}, 
\hat{B}_{\beta}{}^{\dot\nu}\}_{**}
+ \mbox{total derivatives}.
\ee

Summarizing, 
by defining the all-order expressions of the field strengths as
\bea
{\cal H}_{\dot1\dot2}
&=&
H_{\dot1\dot2}+[b_{\dot1}, b_{\dot2}]_{*},
\label{deform-H}
\\
{\cal F}_{\dot1\dot2}
&=&
F_{\dot1\dot2},
\label{deform-Fdot}
\\
\epsilon_{\alpha\beta}{\cal F}^{\beta\dot\mu}
&=&
-\bigg(\partial_{\alpha}
b^{\dot\mu}
-g\{\partial_{\dot\rho}X^{\dot\mu}, 
\hat{B}_{\alpha}{}^{\dot\rho}\}_{**}\bigg),
\label{deform-Fadot}
\\
{\cal F}_{\alpha\beta}
&=&
F_{\alpha\beta}
+g[-F_{\alpha\dot\mu}\hat{B}_{\beta}{}^{\dot\mu}
+F_{\beta\dot\mu}\hat{B}_{\alpha}{}^{\dot\mu}]
+g^2 F_{\dot\mu\dot\nu}
\{\hat{B}_{\alpha}{}^{\dot\mu}, 
\hat{B}_{\beta}{}^{\dot\nu}\}_{**}
+ \mbox{total derivatives}.
\nn \\ &&
\label{deform-Fab}
\eea
We
have shown that the noncommutative gauge theory 
is equivalent to the D3-brane theory 
in large R-R background with the Lagrangian (\ref{action-RR}) to all orders.
The definitions of the field strengths are motivated by the identification 
of the following objects in the dual theories:
\bea
{\cal F}^{\prime}_{\dot1\dot2}&=&
{\cal H}_{\dot1\dot2},
\label{F=H}
\\
\int d^4x\ \frac{1}{2}\epsilon^{\alpha\beta}{\cal F}^{\prime}_{\alpha\beta}F_{\dot1\dot2}&=&
\int d^4x\ \frac{1}{2g}\epsilon^{\alpha\beta}{\cal F}_{\alpha\beta},
\\
\epsilon_{\beta\alpha}\epsilon_{\dot\mu\dot\nu}{\cal F}^{\prime\beta\dot\nu}
&=&
{\cal F}_{\alpha\dot\mu},
\eea
such that formally the Lagrangian (\ref{action-RR}) remains the same.
There is {\em a priori} no reason why this would be sufficient to guarantee
that the field strengths defined above are covariant under gauge transformations.
In fact, 
so far we know nothing about the gauge symmetry of the R-R theory 
beyond the Poisson level.
This is the topic of the next section.


\section{Gauge symmetry to all orders}
\label{5}

In the above we have defined the field strengths (\ref{deform-H})--(\ref{deform-Fab})
for the S-dual theory only by demanding that the dual Lagrangian be expressed 
in an expression which is formally identical to that in the Poisson limit.
This is merely a technical (perhaps technically natural) reason.
We are not guaranteed to have a gauge transformation law 
under which all the field strengths (\ref{deform-H})--(\ref{deform-Fab}) transform covariantly.
The problem is that the operation of electromagnetic duality does not 
preserve the gauge symmetry algebra for non-Abelian gauge theories,
because the Bianchi identity
cannot be stated purely in terms of the field strength 
without explicit reference to the gauge potential.
It is thus not at all obvious how to define 
the gauge transformation laws or the {\em covariant} field strengths at higher orders 
for the dual theory even though their equations of motion can be derived.
In fact, there is always some ambiguity in field redefinitions.
There are infinitely many correct answers 
(although it is still challenging to find any correct answer)
related to each other by field redefinitions,
and we wish to find the simplest answer in compact, elegant expressions. 
Furthermore,
the electromagnetic duality is an equivalence relation that interchanges 
equations of motion with integrability conditions (Bianchi identity).
This means that, at least classically, 
the duality is an equivalence of the spaces of solutions of the two theories.
When we consider off-shell configurations 
(violation of equation of motion in one theory),
we are at the risk of losing the consistency in gauge transformations
(violation of Bianchi identity).
Therefore,
for non-Abelian gauge symmetry,
off-shell gauge symmetry after electromagnetic duality is a subtle problem. 
In fact, the answer that we will find below 
is a gauge symmetry valid for a constrained space of configurations 
for which the space of solutions of the equations of motion is a subspace, 
but not valid for all off-shell configurations.

\subsection{Gauge transformation laws for potentials}

The gauge transformation law for the noncommutative gauge theory is well known.
The gauge transformation law of the potential is
\bea
\delta_{\Lambda}a^{\prime}_{A}=\partial_{A}\Lambda-[\Lambda, a^{\prime}_{A}]_{*},
\eea
which implies that the field strength defined by
\bea
{\cal F}^{\prime}_{AB}=F^{\prime}_{AB}+[a^{\prime}_{A}, a^{\prime}_{B}]_{*}
\eea
is covariant.
That is
\bea
\delta_{\Lambda}{\cal F}^{\prime}_{AB}=-[\Lambda, {\cal F}^{\prime}_{AB}]_*.
\eea

A closer examination of the derivation of S-duality for the D3-brane 
gives us hints about the gauge transformation laws in the R-R theory.
With some trial and errors,
we 
find 
that the gauge transformation laws are given by
\bea
\delta_{\Lambda}b^{\dot\mu}
&=&\kappa^{\dot\mu}
+g
\{\kappa^{\dot\nu}, 
\partial_{\dot\nu}
b^{\dot\mu}\}_{**},
\label{gt-b}
\\
\delta_{\Lambda}a_{\dot\mu}
&=&\partial_{\alpha}\lambda
+g\bigg(
\{\kappa^{\dot\nu}, 
\partial_{\dot\nu}
a_{\dm}\}_{**}
+\{a_{\dot\nu}, 
\partial_{\dm}
\kappa^{\dot\nu}\}_{**}\bigg),
\label{gt-adm}
\\
\delta_{\Lambda}\hat{B}_{\alpha}{}^{\dot\mu}
&=&\partial_{\alpha}\kappa^{\dot\mu}
+g
\bigg(
\{\kappa^{\dot\nu}, 
\partial_{\dot\nu}
\hat{B}_{\alpha}{}^{\dot\mu}\}_{**}
-\{
\hat{B}_{\alpha}{}^{\dot\nu}, 
\partial_{\dot\nu}
\kappa^{\dot\mu}\}_{**}\bigg).
\label{gt-B}
\eea
The function $\lambda$ is the gauge parameter of the usual $U(1)$ gauge symmetry
independent of $\kappa^{\dm}$,
while $\kappa^{\dm}$ is the gauge parameter for 
the area-preserving diffeomorphism on the subspace of $x^{\dot1}, x^{\dot2}$.
The gauge transformation law for $a_{\alpha}$ is not given here 
because ${\cal F}_{\alpha\beta}$ is only determined up to total derivatives.
In principle it is possible to determined ${\cal F}_{\alpha\beta}$ to all orders 
by the requirement of covariance, 
and then to determine the gauge transformation of $a_{\alpha}$ accordingly.
We leave this problem for future investigation.

The guess of the all-order form of the gauge transformation of $a_A$ is 
a natural extension of the corresponding expression in the Poisson limit. 
The gauge transformations of the rest of the potentials are motivated by the demand that, 
to all orders,
\bea
b^{\dot\mu}&=&
\epsilon^{\dot\mu\dot\nu}a_{\dot\nu}^{\prime},
\\
\hat{B}_{\alpha}{}^{\dot\mu}&=&
\epsilon^{\dot\mu\dot\nu}\partial_{\dot\nu}a^{\prime}_{\alpha}.
\eea
We have to emphasize that the on-shell condition 
$\partial_{\dot\mu}\hat{B}_{\alpha}{}^{\dot\mu}=0$ 
is needed for the conjecture to be consistent.
This condition is the equation of motion of $a_{\alpha}$.

One should verify that the field strengths are covariant and 
the algebra of gauge transformation is closed.
These would be non-trivial consistency checks of our conjecture.
According to the transformation laws defined above,
except ${\cal F}_{\alpha\beta}$ which is only defined up to total derivatives,
the field strengths are indeed covariant:
\bea
\delta_{\Lambda}{\cal H}_{\dot1\dot2}
&=&
g\{\kappa^{\dot\nu}, \partial_{\dot\nu}{\cal H}_{\dot1\dot2}\}_{**},
\\
\delta_{\Lambda}{\cal F}_{\dot1\dot2}
&=&
g\{\kappa^{\dot\nu}, \partial_{\dot\nu}{\cal F}_{\dot1\dot2}\}_{**},
\\
\delta_{\Lambda}{\cal F}_{\alpha\dot\mu}
&=&
g\{\kappa^{\dot\nu}, \partial_{\dot\nu}{\cal F}_{\alpha\dot\mu}\}_{**},
\eea
In the next subsection we show that 
the gauge transformation algebra is closed.

\subsection{Gauge algebra and closedness}

The product $**$ has two important algebraic properties 
that will be useful for our calculation.
First, it is associative under integration:
\bea
\int{}\{A, \{B, C\}_{**}\}_{**}=\int{}\{\{A, B\}_{**}, C\}_{**},
\eea
although it is not associative without integration.
The second property is that
\bea
g\epsilon^{\dot\mu\dot\nu}\{\partial_{\dot\mu}f, \partial_{\dot\nu}g\}_{**}=[f, g]_*.
\eea
As a result of the second property,
one can rewrite the gauge transformation law as
\bea
\delta_{\Lambda}(\mbox{field strength})
=g\{\kappa^{\dot\nu}, \partial_{\dot\nu}(\mbox{field strength})\}_{**}
=-\lbrack\Lambda, (\mbox{field strength})\rbrack_*,
\eea
where $\kappa$ and $\Lambda$ are related by
\be
\kappa^{\dot\nu}=\epsilon^{\dot\nu\dot\rho}\partial_{\dot\rho}\Lambda. 
\ee
Similarly,
one can rewrite the gauge transformation laws (\ref{gt-b})--(\ref{gt-B})
of the gauge potentials as
\bea
\delta_{\Lambda}b^{\dot\mu}&=&\epsilon^{\dot\mu\dot\nu}\partial_{\dot\nu}\Lambda
-[\Lambda, b^{\dot\mu}]_{*},
\label{gt-b1}
\\
\delta_{\Lambda} a_{\dot\mu}
&=&\partial_{\dot\mu}\lambda-[\Lambda, a_{\dot\mu}]_* 
+g\epsilon^{\dot\nu\dot\rho}
\{\partial_{\dm}\partial_{\dot\rho}\Lambda, a_{\dot\nu}\}_{**},
\label{gt-adm1}
\\
\delta_{\Lambda}\hat{B}_{\alpha}{}^{\dot\mu}&=&
\epsilon^{\dot\mu\dot\nu}\partial_{\dot\nu}\partial_{\alpha}\Lambda-
[\Lambda, \hat{B}_{\alpha}{}^{\dot\mu}]_*
-g\epsilon^{\dot\mu\dot\nu}\{\partial_{\dot\nu}\partial_{\dot\rho}
\Lambda, \hat{B}_{\alpha}{}^{\dot\rho}\}_{**}.
\label{gt-B1}
\eea
The function $\lambda$ is the gauge parameter of the usual $U(1)$ gauge symmetry
independent of $\Lambda$.

The closedness of the algebra of gauge transformation is 
proven by the existence of the gauge parameters 
$\Lambda_3$ and $\lambda_3$ such that, 
for arbitrary gauge transformation $\delta_1, \delta_2$ 
with the parameters $(\Lambda_1, \lambda_1)$, $(\Lambda_2, \lambda_2)$, 
we have a gauge transformation $\d_3$ with parameters $(\Lambda_3, \lambda_3)$
satisfying the closedness relation
\be
[\d_1, \d_2] = \d_3
\label{d1d2d3}
\ee
for all gauge potentials.
It is remarkable that such a solution exists:
\bea
\Lambda_3&=&g\epsilon_{\dot\rho\dot\sigma}
\{\kappa_1{}^{\dot\rho}, \kappa_2{}^{\dot\sigma}\}_{**},
\label{Lambda3}
\\
\lambda_3&=&
F^{\dot\mu}(\partial_x, \partial_y, \partial_z)\Lambda_1(x)\Lambda_2(y)
a_{\dot\mu}(z)\mid_{y=x,z=x}.
\label{lambda3}
\eea
Here $F^{\dot\mu}$ is a tri-linear operator acting on the gauge parameters 
$\Lambda_1, \Lambda_2$ and $a_{\mu}$.
It is
\bea
F^{\dot\mu}(ik, ik', ip) 
&\equiv&
i\frac{(k\times k')}{(k+k')\times p}
\bigg(
\gnum{k\times p}\gnum{k'\times(k+p)}
-\gnum{k'\times p}\gnum{k\times(k'+p)}
\bigg) \cdot
\nn \\
&&
\frac{({k+k^{\prime}+p})^{\dot\rho}}{(k+k^{\prime}+p)^2}
\lbrack k_{\dot\rho} k^{\prime}\times(k+k^{\prime}+p)
-k^{\prime}_{\dot\rho} k\times(k+k^{\prime}+p)]
\rbrack p_{\dot\nu}\epsilon^{\dot\nu\dot\mu},
\eea
where
\bea
k\times k' 
&\equiv& 
\epsilon^{\dot\mu\dot\nu}k_{\dot\mu}k'_{\dot\nu},
\\
\gnum{s} 
&\equiv&
\frac{e^{gs/2} - e^{-gs/2}}{s}.
\eea

\subsection{Equations of motion}

As we have proven in Sec.\ref{4},
in terms of the deformed gauge symmetry and field strengths,
the Lagrangian of the R-R theory has a natural extension to higher orders 
given by the same form (\ref{action-RR})
\bea
{\cal L}_{RR} \equiv 
-\frac{1}{2}{\cal H}^2_{\dot1\dot2}+\frac{1}{2}{\cal F}_{\alpha\dot\mu}{\cal F}^{\alpha\dot\mu}
-\frac{1}{4}F_{\dot\mu\dot\nu}F^{\dot\mu\dot\nu}
+\frac{1}{2g}\epsilon^{\alpha\beta}{\cal F}_{\alpha\beta},
\eea
but with the field strengths defined by (\ref{deform-H})--(\ref{deform-Fab}).

The equations of motion of the deformed R-R theory to all orders 
derived by varying the fields 
$b^{\dot{\mu}}, a_{\alpha}, a_{\dot\mu}, \hat{B}_{\alpha}{}^{\dot\mu}$ are, respectively,
\bea
(D_{\dot\mu}{\cal H}_{\dot1\dot2})_*
+ \epsilon_{\alpha\beta}(D^{\alpha}{\cal F}^{\beta}{}_{\dot\mu})_{**}&=&0,
\\
\partial_{\dot\mu}\hat{B}_{\alpha}{}^{\dot\mu}&=&0,
\\
\partial_{\dot\mu}\bigg(F^{\dot\mu\dot\nu}
-g\epsilon^{\alpha\beta}\{\hat{B}_{\alpha}{}^{\dot\mu}, \hat{B}_{\beta}{}^{\dot\nu}\}_{**}\bigg)
+\epsilon^{\alpha\beta}\partial_{\alpha}\hat{B}_{\beta}{}^{\dot\nu}&=&0,
\\
\{V_{\dot\mu}{}^{\dot\nu}, \bigg(\partial^{\alpha}b_{\dot\nu}
-\{V^{\dot\rho}{}_{\dot\nu}, \hat{B}^{\alpha}{}_{\dot\rho}\}_{**}\bigg)\}_{**}
+\epsilon^{\alpha\beta}F_{\beta\dot\mu}+g\epsilon^{\alpha\beta}
\{F_{\dot\mu\dot\nu},
\hat{B}_{\beta}{}^{\dot\nu}\}_{**}&=&0,
\label{eom-B}
\eea
where we define the covariant derivatives as
\bea
(D_{\alpha}\Phi)_{**}&\equiv&
\partial_{\alpha}\Phi-g\{\hat{B}_{\alpha}{}^{\dot\rho}, \partial_{\dot\rho} \Phi\}_{**},
\label{cov-D-a}
\\
(D_{\dot\mu}\Phi)_*&\equiv&
\partial_{\dot\mu}\Phi-\epsilon_{\dot\mu\dot\rho}[b^{\dot\rho}, \Phi]_*.
\label{cov-D-dm}
\eea

The deformed NS-NS theory is just the well-known noncommutative gauge theory.
The equations of motion are
\bea
(D'_{\dot\mu}{\cal F}^{\prime\dot\mu\dot\nu})_*
+(D'_{\alpha}{\cal F}^{\prime\alpha\dot\nu})_*&=&0,
\\
\epsilon^{\alpha\beta}(D'_{\beta}{\cal F}^{\prime}_{01})_*
-(D'_{\dot\mu}{\cal F}^{\prime\dot\mu\alpha})_*&=&0,
\eea 
where
\bea
(D'_{\alpha}\Phi)_*&=&
\partial_{\alpha}\Phi+\lbrack a^{\prime}_{\alpha}, \Phi\rbrack_*,
\\
(D'_{\dot\mu}\Phi)_*&=&
\partial_{\dot\mu}\Phi+\lbrack a^{\prime}_{\dot\mu}, \Phi\rbrack_*.
\eea
One can check that the equations of motion of the NS-NS and R-R theories 
are equivalent.


\section{Discussion}
\label{6}

With hints drawn from the NP M5-brane theory 
for an M5-brane in large $C$-field background,
we derived the effective Lagrangian (\ref{action-RR})
for a D3-brane in R-R background 
from its S-dual configuration -- 
the noncommutative $U(1)$ gauge theory for a D3-brane 
in constant NS-NS $B$-field background.
We have also constructed the gauge symmetry 
for the D3-brane in the R-R background, 
including covariant field strengths (\ref{deform-H})--(\ref{deform-Fab})
and covariant derivatives (\ref{cov-D-a})--(\ref{cov-D-dm}).
The gauge transformation laws of the potentials (\ref{gt-b})--(\ref{gt-B})
are defined in terms of a non-associative product (\ref{non-assoc-product})
involving infinite derivatives.
The gauge symmetry algebra (\ref{d1d2d3})--(\ref{lambda3}) 
of the D3-brane in R-R background is a mixture of 
the usual $U(1)$ gauge symmetry and a deformation of area-preserving diffeomorphism.

According to \cite{Ganor:2000my,Rey:2000hh},
the dual theory of a noncommutative gauge theory 
is again a noncommutative gauge theory
at the leading order (Poisson level).
This result is puzzling \cite{Rey:2000hh} because, 
from the viewpoint of string theory, 
we do not expect world-volume noncommutativity for the dual D3-brane 
according to how a fundamental string is coupled to the R-R potential.
This puzzle is resolved by our result of the dual theory,
in which the $U(1)$ gauge symmetry is commutative,
but there is an additional gauge symmetry
(that of deformed area-preserving diffeomorphism).
Integrating out the gauge potential of this additional gauge symmetry 
deforms the $U(1)$ symmetry so that it looks like 
noncommutative $U(1)$ gauge symmetry at the leading order, 
but it is merely a coincidence. 

In the NS-NS $B$-field background,
open strings ending on the D3-brane 
interact continuously with the $B$-field 
so that the scattering of open string oscillation modes is modified. 
From the perspective of the D3-brane effective theory,
the effect of this modification (in a certain limit) 
is equivalent to turning on the noncommutativity of world-volume coordinates
\cite{Chu-Ho,Schomerus:1999ug,Seiberg:1999vs}.
The corresponding S-dual picture of the above 
has open D1-branes ending on D3-branes in the R-R background.
Although it is less emphasized in the literature, 
an immediate implication of S-duality is
that the world-volume theory of a D3-brane 
can be viewed as the field theory of 
massless oscillation modes of open D1-branes.
The contributions of D1-branes and open strings do not add up 
to give twice the field content.
Instead, they are equivalent descriptions of the same degrees of freedom 
related to each other by electromagnetic duality.

In the R-R theory,
the gauge potential $b^{\dm}$ arises as a massless oscillation mode of an open D1-brane,
and is in charge of the (deformed) area-preserving diffeomorphism.
According to (\ref{Sgauge10Dp}), 
the equation of motion at the lowest order indicates that 
$b^{\dm}$ shares the same physical degrees of freedom with 
the gauge potential $a_{\alpha}$ of the usual $U(1)$ symmetry \cite{Ho:2013paa}.
In principle, 
one can integrate out $b^{\dm}$, 
at least perturbatively, 
but it is much more convenient to keep $b^{\dm}$ in the Lagrangian 
so that the covariant derivatives 
with respect to the area-preserving diffeomorphism can be used 
to concisely describe interactions.
This is one of the crucial facts that has made it possible 
to write down the Lagrangian of the R-R theory explicitly.
Another crucial fact utilized in the duality transformation is that 
the Bianchi identity for the field strength $F_{\dot1\dot2}$ in 2 dimensions is trivial.

Some readers may wonder why the parameter $g$ is not inverted by 
the duality transformation.
We explained in Sec.\ref{coupling-inv} that, 
even though this is still true at the leading order 
with a suitable choice of field variables,
it is more convenient to choose the field variables differently 
in order to incorporate all-order effects in a compact way, 
at least for the pure gauge theory.
From the viewpoint of string theory, 
although S-duality transforms the closed-string coupling $g_s$
to its inverse $g'_s = 1/g_s$,
the gauge coupling constant of the D3-brane world-volume theory
is not the same as the closed-string coupling.
For the NS-NS theory in the Seiberg-Witten limit, 
the gauge coupling constant is
\be
\frac{1}{g_{YM}^2} \simeq \frac{(\det g)^{1/2}}{2\pi\alpha' g_s}\frac{1}{|B_{\dot1\dot2}|},
\label{gYM}
\ee
where $(\det g)$ is the determinant of the closed-string metric.
The combination of the first fractional factor has a fixed limit in the Seiberg-Witten limit,
as well as $B_{\dot1\dot2}$.
If we had considered the field theory of open string oscillation modes for the R-R theory,
we would expect the factor $g_s$ in $g_{YM}$ inverted, 
and the gauge coupling would diverge in the limit.
(Even though $\alpha'$ is also changed by S-duality,
it does not change the conclusion.)
However, as we mentioned above, 
we are using open D1-brane oscillation modes represented by $b_{\dm}$ 
in the R-R theory,
we expect that the gauge coupling to be given by a formula different from (\ref{gYM}).
What we saw was that the gauge coupling was not inverted in the R-R theory.

The experience we have learned in this project 
may help us to understand more about electromagnetic duality in non-Abelian gauge theories.
It will be interesting to extend the D3-brane theory for the R-R background 
to a system of multiple D3-branes.
It will also be desirable to uplift the D3-brane theory in R-R background 
to D4-brane in R-R background via T-duality,
so that the gauge symmetry of the area-preserving diffeomorphism 
is lifted to a volume-preserving diffeomorphism, 
and to see how the no-go theorem \cite{Chen:2010ny} will be circumvented.

\section*{Acknowledgement}

We thank Heng-Yu Chen, Wei-Ming Chen, 
Chong-Sun Chu, Takeo Inami, Branislav Jur\v{c}o,
Hsien-chung Kao, Pei-Wen Peggy Kao, 
Yutaka Matsuo, Hiroaki Nakajima, 
Peter Schupp and Tomohisa Takimi for discussions.
This work is supported in part by 
NTU (grant \#NTU-CDP-102R7708), 
and by National Science Council, Taiwan, R.O.C.

\end{document}